\documentclass[CRPHYS,Unicode,manuscript]{cedram}

\title{Mathematical physics of dilute Bose gases}
\alttitle{Physique mathématique des gaz de bosons dilués}

\author{\firstname{Jan Philip} \lastname{Solovej}}
\address{QMATH, Department of Mathematical Sciences\\University of Copenhagen\\Universitetsparken 5, DK-2100, Copenhagen, DENMARK}
\email{solovej@math.ku.dk}
\thanks{The work is supported by the VILLUM Foundation grant \#10059.}
\ESM{There is no supplementary material for this article.}
\keywords{Ground states of Bose gases, Cold atomic gases, Bose-Einstein Condensation, Bogolyubov approximation, Lee-Huang-Yang formula}
\altkeywords{États fondamentaux des gaz de bosons, gaz atomiques froids, condensation de Bose-Einstein, approximation de Bogolioubov, formule de Lee-Huang-Yang}

\begin{abstract}
  We discuss recent progress in the mathematical analysis of dilute Bose gases. We review results in one to three dimensions, but the focus will be on three dimensions. In all dimensions we have a two term asymptotic expansion of the ground state energy density by an expression that depends only on the scattering length of the potential. In dimension three this is the celebrated Lee-Huang-Yang formula.  In dimensions two and three the dilute limit is a weakly interacting regime whereas in dimension one it is rather strongly interacting. We sketch briefly the mathematical difficulties and review some remaining open problems in the field.   
\end{abstract}

\begin{altabstract}
Nous discutons des progrès récents dans l'analyse mathématique des gaz de bosons dilués. Nous passons en revue les résultats obtenus en une, deux et trois dimensions, mais l'accent sera mis sur le cas tridimensionnel. En chaque dimension, nous disposons d'un développement asymptotique à deux termes de la densité d'énergie de l'état fondamental, sous la forme d'expressions ne dépendant que de la longueur de diffusion du potentiel. En dimension trois, il s'agit de la célèbre formule de Lee-Huang-Yang.  En dimension deux et en dimension trois, la limite diluée est dans le régime d'interaction faible, alors qu'en dimension un, elle est au contraire en interaction forte. Nous esquissons brièvement les difficultés mathématiques rencontrées et passons en revue quelques problèmes encore ouverts dans ce domaine.
\end{altabstract}

\begin{document}

% Use the \maketitle command after the abstract
\maketitle

% Example of section

% Example of subsection
\section{Introduction}
Our goal here is to discuss the mathematical analysis of dilute Bose gases and, in particular, the ground state energy density in this limit.  We consider $N$ bosons moving in a cube $[0,L]^d$ of length $L>0$ in dimension $d=1,2$ or $3$. We use units in which $\hbar=1$ and the mass of the boson is $1/2$. The kinetic energy of a boson is thus represented by the Laplace operator $-\Delta=-\nabla^2$. The bosons will  interact through a two-body potential $v$ such that the system Hamiltonian is given by
\begin{equation}\label{eq:Ham}
    H_N(v)=\sum_{i=1}^N-\Delta_i+\sum_{i<j}v(x_i-x_j)
\end{equation}
acting on an appropriate subspace ${\mathcal D}$ of symmetric square integrable functions on $[0,L]^{Nd}$, i.e., of $L^2([0,L]^{Nd})=\bigotimes^NL^2([0,L]^{d})$. Symmetry  refers to interchange of the $N$ boson coordinates in $[0,L]^{d}$. The problem will not depend very much on the choice of the domain ${\mathcal D}$, i.e., boundary conditions (e.g., Dirichlet, Neumann or periodic) will, in general, not be important. We will nevertheless consider it to consist of functions that have two continuous partial derivatives and that satisfy periodic boundary conditions on $[0,L]^{d}$. This is of course equivalent to consider particles on a torus. 

The assumptions on the two-body potential $v$ are, however, important for the known mathematical results. These assumptions relate to some of the open problems we will highlight. We need to assume that $v\geq 0$ and that $v$ is spherically symmetric. For simplicity here we will also assume that $v$ vanishes outside a bounded region. This last assumption can certainly be relaxed to some sufficiently fast decay, but the assumption that $v$ is non-negative is important and removing it is a challenge. The only known work, relating to what we discuss,  allowing $v$ to be slightly negative is \cite{Yin-negV}. 
The potential is allowed to be rather singular. The most important example is maybe the hard-core potential
\begin{align}\label{eq:hc}
      v(x)=\left\{\begin{aligned} \infty,& &|x|<R\\ 
      0,&&|x|>R
\end{aligned}\right..
\end{align}

We will here be concerned with the {\it ground state energy}, i.e., the case of zero temperature $T=0$:
$$
E_L(N,v)=\inf_{\Psi\in {\mathcal D}}\frac{\langle\Psi| H_N|\Psi\rangle}{\langle\Psi|\Psi\rangle}
$$
The inf here is not necessarily a min, because the normalized ground state $\Psi_0$ which satisfies
$$
E_L(N,v)=\langle\Psi_0| H_N|\Psi_0\rangle
$$
may not be in the domain ${\mathcal D}$ that we chose, but may only be approximated from ${\mathcal D}$. Moreover here it is not really important that ${\mathcal D}$ consists of symmetric functions. If we ignored this condition the infimum would be the same, i.e., the ground state $\Psi_0$ is automatically symmetric, i.e., bosonic.

We are really interested in the {\it thermodynamic limit} of the ground state energy density
$$
e(\rho,v)=\lim_{\substack{L\to\infty\\ NL^{-d}\to \rho}} L^{-d}E_L(N)
$$
for a particle number density $\rho\geq0$. It is not difficult to prove that this limit exists and that $e(\rho)$ is a convex function of $\rho$. 

While the energy $E_L(N,v)$ may depend on the boundary conditions we choose on the boundary of $[0,L]^d$ the thermodynamic energy density $e(\rho)$ will not. 

The goal is to understand the structure of the ground state and of $e(\rho)$ in the dilute limit $\rho\to0$. In particular we would like to prove Bose-Einstein Condensation (BEC) in $d=2$ and $3$ and to understand the asymptotic expansion of $e(\rho)$ in the dilute limit. 

We should, however, first understand what we mean exactly by the dilute limit, i.e., what is the right parameter to compare $\rho$ to. For that we need to introduce the {\it Scattering Length} of the potential $v$. We will do that in Section~\ref{sec:scatt}. In Section~\ref{sec:mainresult} we formulate the main mathematical results on the dilute limit of the ground state energy density. In section~\ref{sec:2nd} we review the Bogolubov approximation as a variational theory and explain what it gives in the dilute limit. In Section~\ref{sec:BEC} we discuss Bose-Einstein condensation and what is known about it rigorously.
 In Section~\ref{sec:proof} we sketch some of the ideas involved in proving the energy asymptotics in three dimensions---The Lee-Huang-Yang formula. Finally, in Section~\ref{sec:1D} we make a few remarks about the one-dimensional case 
 where we do not have BEC and compare the result on the dilute limit for  general potentials with the exact results for the hard-core gas and the Lieb-Liniger gas. We conclude in Section~\ref{sec:conclusion} with a summary and open questions. 

\section{Scattering length}\label{sec:scatt}
To introduce the scattering length we consider the zero energy scattering equation \cite{bosebook}
$$
-\Delta u+\frac12 vu=0,\quad \text{on }{{\mathbb R}^d}.
$$
Here the factor $1/2$ may be seen as coming from the reduced mass of the two-body problem, i.e., that there is a factor 2 in front of the Laplacian in relative coordinates. 

If $v$ vanishes outside a ball of radius $R$ there will in dimension $d=1,2,3$ be a unique solution $u$ satisfying that for  $|x|>R$ 
$$
d=1:\ u(x)=|x|-a,\qquad d=2:\ u(x)=\ln(|x|/a),\qquad d=3:\ u(x)=1-\frac{a}{|x|}
$$
for a unique length scale $a$ called the {\it scattering length} of the potential $v$. 

Note that we must have $a\leq R$ otherwise the scattering function $u$ changes sign and hence, by the Sturm oscillation principle, the potential $v$ will have a two-body bound state. For the hard-core potential \eqref{eq:hc}  we have $a=R$.

Under our assumption $v\geq0$ this cannot happen  and moreover in dimensions $2$ and $3$, $a\geq0$. In dimension $1$ we may have $a$ negative even if $v$ is positive, e.g., for the delta interaction, i.e., the Lieb-Liniger gas.

The free case $v=0$ corresponds to $a=0$ in dimensions $d=2$ and $3$, but in dimension $d=1$ it corresponds to $a=-\infty$.  Since $a=0$ for the free gas in $2$ and $3$ dimensions the dilute limit can also be seen as the weakly interacting case. In dimension $1$, however, the dilute gas corresponds rather to the strongly interacting case. 

In dimension three we have $0\leq u\leq 1$ and it will be convenient for us to write the scattering solution as 
$
u(x)=1-\omega(x)
$,
with $0\leq\omega(x)\leq 1$ and $\omega(x)=a/|x|$ for large $|x|$. It is also easy to see, in this case, that 
\begin{equation}\label{eq:u2a}
8\pi a=\int v(x)u(x)dx \leq \int v(x)dx. 
\end{equation}

Equivalently the scattering length could have been defined from the two particle ground state energy satisfying the following $L\to\infty$ asymptotics
\begin{align}
    E_L(N=2,v)=\begin{cases}
        2\pi^2 (L-a)^{-2}+o(aL^{-3}), & d=1\\
        4\pi L^{-2}|\ln\frac{a}{L}|^{-1}(1+o(1)),&d=2\\
        8\pi aL^{-3}(1+o(1)),&d=3
    \end{cases}.
\end{align}
Here it is important that we consider wave functions satisfying periodic boundary conditions. 

Having defined the scattering length we can make the condition of diluteness more precise. Indeed, it refers to the dimensionless parameter  $\rho|a|^d$ being small.

\section{Main result on energy density asymptotics}\label{sec:mainresult}
The main mathematical results on the ground state energy density of dilute gases is collected in the following theorem. 

\begin{theorem}[Ground state energy density asymptotics]\label{thm:main} For $v\geq 0$  with finite range in the ball $|x|\leq R$ we find in the limit  $\rho|a|^d\to 0$ that
\begin{align*}
 \text{d=3:}&&e(\rho)=&4\pi\rho^2 a\left(1+\frac{128}{15\sqrt{\pi}} (\rho a^3)^{1/2}+o((\rho a^3)^{1/2} )\right)\\
 \text{d=2:}&&e(\rho)=&4\pi\rho^2 Y\left(1+Y\log(Y\pi)+(2\Gamma+\frac12)Y+o(Y)\right),&Y=|\log(\rho a^2)|^{-1}\\
 \text{d=1:}&&e(\rho)=&\frac{\pi^2}3\rho^3(1+2\rho a+o(\rho R)).
\end{align*}
 The result in dimension three requires $v\in L^3$, i.e., that the $|v|^3$ is integrable, but this is only required to prove the upper bound in the estimate. Above $\Gamma $ is the Euler–Mascheroni constant.
\end{theorem}
The error in dimension one is not only in terms of $\rho a$ but involves the range $R$. In the one-dimensional case we may have $a=0$ for a non-trivial potential and we can then not write the error solely in terms of $\rho a$. 

 The leading term in the asymptotic formula in dimension 3 goes back to Lenz \cite{Lenz1929}. The full two-term asymptotics in $d=3$, {\it the Lee-Huang-Yang formula},  goes back to the 1957 work of Lee-Huang-Yang  \cite{Lee1957} based on an analysis of the hard-core gas. They conjectured that the two terms should hold universally. The universality of the upper bound in the leading term was established by Dyson \cite{dyson} in 1957 and the leading order lower bound by Lieb and Yngvason in 1998 \cite{LiebYngvason1998}. The upper bound to the Lee-Huang-Yang precision was proved by Yau and Yin \cite{YauYin} and simplified in  
 \cite{bastietal21}. It requires, as explained in the theorem, the potential $v$ to satisfy integrability conditions and does not include the hard-core case. The lower bound was proved in \cite{Fournais2020,Fournais2022}. It holds also for the hard-core potential. It is curious that Lee, Huang, and Yang derived the formula from an analysis of the hard-core gas while this case is still a mathematical challenge. The recent paper \cite{Bastietal24} derives an upper bound on the ground state energy density for the hard-core gas with an error of the order of the Lee-Huang-Yang correction but with too large a constant. In \cite{bastietal23} a version of the upper bound to the Lee-Huang-Yang precision was derived in a confined Gross-Pitaevskii regime. The Lee-Huang-Yang formula was observed experimentally in a cold dilute gas of  ${}^7$Li atoms in \cite{Navon2011, Navon2010}.
 
 The two-dimensional leading-order case was studied by Schick in \cite{Schick1971} and rigorously by Lieb and Yngvason in \cite{LY2D}. The higher corrections were studied in \cite{Hines1978,Cherny2001,Yang2008, Mora2009}. A complete proof including the hard-core case for both the upper and lower bounds was given in \cite{Fournais2024}. The one-dimensional case was proved in \cite{agerskov2024}.

 In dimension three corrections beyond the Lee-Huang-Yang order were predicted by T.T.\ Wu \cite{Wu1959} to be of the form
 \begin{equation}\label{eq:Wu}
e(\rho) = 4\pi\rho^2 a \left(1+\frac{128}{15\pi^{1/2}}(\rho a^3)^{1/2} +8(\frac{4\pi}{3}- \sqrt{3})\rho a^3 \ln(\rho a^3) +{\mathcal E}\rho a^3 + o(\rho a^3)\right).
\end{equation}
The constant ${\mathcal E}$ can be found more explicitly in \cite{Tan2008} which claims 
$$
{\mathcal E}=\frac{D}{12\pi a^4} + \pi \frac{r_s}{a} + C,
$$
where $C$ is a universal constant, $D$ is the three-body scattering hyper-volume, and $r_s$ is the effective range (for the hard-core $r_s=2a/3$, see \cite{Tan2008}). From the rigorous point of view, all terms beyond Lee-Huang-Yang are a challenge for thermodynamic systems. In a confined setting progress on the third term above was achieved in \cite{caraci2024ordercorrectionsgroundstate}.

In dimension two Mora and Castin \cite{Mora2009} find\footnote{I thank the referee for suggesting to add the formula in this form which has not appeared previously.}
\begin{equation}\label{eq:Moracastin}
e(\rho)=4\pi\rho^2 Y \left(1+ Y \big(\log( Y\pi) +2\Gamma+1/2\big) + Y^2 \big(\log( Y\pi) +2\Gamma+1\big)^2- \frac{8 I}{\pi}Y^2 + o(Y^2)\right),
\end{equation}
where $I$ is a universal constant.

\section{Second quantization and Bogolyubov Theory}\label{sec:2nd}

In this section we will briefly sketch how to understand the Lee-Huang-Yang formula from the Bogolyubov approximation \cite{Bogolyubov1947}. We shall however proceed in a somewhat non-standard way. Usually the Bogolyubov approximation approximates the Hamiltonian by a quadratic Hamiltonian in bosonic creation and annihilation operators. Such a Hamiltonian will be minimized by a Gaussian or quasi-free state. Rather than approximating the Hamiltonian we will introduce Bogolyubov theory as a variational approach where we leave the Hamiltonian unchanged but restrict to quasi-free states. For details see \cite{NapiorkowskiReuversI, NApiorkowskiReuversII}. This has the advantage that the resulting approximation is actually an upper bond on the true energy density. 

We write the Hamiltonian (\ref{eq:Ham}) on the 3-dimensional torus in second quantized form. 
$$
H_N(v)=\sum_{p\in (2\pi {\mathbb Z}/L)^3}
p^2a_p^\dagger a_p+\frac1{2L^3}\sum_{p,q,k}\widehat{v}(k)a_{p+k}^\dagger a_{q-k}^\dagger a_q a_p.
$$
Here we use the convention $\widehat{v}(p)=\int v(x)e^{-ipx}dx$ for the Fourier transform, i.e., on the torus the interaction $v$ between two particles is a function of the distance between them measured on the torus.  

We perform a unitary transformation $a_0\to a_0+\sqrt{\rho_0 L^3}$ that rewrites the Hamiltonian in terms of variables expanding around the condensate with density $\rho_0$ in the zero momentum state. 

After this we evaluate in a quasi-free translation invariant state where all expectation values can be expressed by Wick's Theorem in terms of the two quantities: 
$$
\gamma(p)=\langle a^\dagger_pa_p\rangle,\quad 
\alpha(p)=\langle a^\dagger_pa^\dagger_{-p}\rangle. 
$$
It is sufficient here to restrict to $\alpha(p)$ real. They satisfy that the matrix 
$$
\begin{pmatrix}
    \gamma(p)&\alpha(p)\\ \alpha(p)&1+\gamma(p)
\end{pmatrix}
$$
is positive semi-definite and hence 
$$
\alpha(p)^2\leq \gamma(p)(1+\gamma(p)).
$$
We find the energy density in the thermodynamic limit expressed as the functional
\begin{align}
{\mathcal E}_{\rm Bog}(\rho_0,\gamma,\alpha)= &
(2\pi)^{-3}\int p^2 \gamma(p) dp+
\frac12\widehat{v}(0)\left(\rho_0+(2\pi)^{-3}\int\gamma(p)dp\right)^2\nonumber\\&
+\rho_0(2\pi)^{-3}\int \widehat{v}(p)(\gamma(p)+\alpha(p))dp \nonumber\\&+
\frac12(2\pi)^{-6}\iint \widehat{v}(p-q)(\gamma(p)\gamma(q)+\alpha(p)\alpha(q))dpdq.\label{eq:bofunctional}
\end{align}

We will now sketch how to get the leading term in the $d=3$ expansion from this expression. The minimizer will have $\alpha(p)=-\sqrt{\gamma(p)(1+\gamma(p))}$. It can be shown that the minimizer of the Bogolyubov functional (\ref{eq:bofunctional}) has almost complete condensation, i.e., 
$$
\rho:=\rho_0+(2\pi)^{-3}\int\gamma(p) dp\approx\rho_0
$$
and we will have $\gamma(p)\ll1$. We can therefore approximate $\gamma(p)\approx \alpha(p)^2$. If we write the inverse Fourier transform of $\alpha$ as $\check\alpha(x)$, we see that 
$$
(2\pi)^{-3}\int p^2\gamma(p)dp\approx \int |\nabla\check\alpha(x)|^2dx.
$$
The energy density is therefore approximated by 
\begin{equation}\label{eq:alphaenapprox}
\int |\nabla\check\alpha(x)|^2dx +\rho\int v(x)\check{\alpha}(x)dx+\frac12\int v(x)|\check{\alpha}(x)|^2dx+\frac12\widehat{v}(0)\rho^2.
\end{equation}
With these approximations the variational equation for $\check\alpha$ is 
\begin{equation}\label{eq:alphascat}
-\Delta\check\alpha+\frac12v\check\alpha+\frac12\rho v=0
\end{equation}
We can rewrite this equation as 
$$
-\Delta(1+\rho^{-1}\check\alpha)+\frac12v(1+\rho^{-1}\check\alpha)=0.
$$
we recognize this as the scattering equation and conclude that $\check\alpha(x)\approx -\rho \omega(x)$. 
Multiplying equation (\ref{eq:alphascat}) by $\check\alpha$ and integrating we find that the approximation (\ref{eq:alphaenapprox}) to the energy density gives 
$$\frac12 \rho\int v(x)\check\alpha(x)dx +\frac12\widehat{v}(0)\rho^2=\frac12\rho^2\int v(x) (1-\omega(x))dx=4\pi\rho^2 a,
$$
i.e., exactly the leading term in the 3-dimensional energy asymptotics.
Here we used that the scattering solution is $u(x)=1-\omega(x)$ and the identity (\ref{eq:u2a}). 

In the usual formulation of the Bogolyubov approximation  $a_0$ is replaced by a number (c-number substitution) and the remaining Hamiltonian is approximated by an operator quadratic in the remaining creation and annihilation operators by ignoring cubic and quartic terms.  Proceeding in such a fashion it is, actually not even possible to get the correct leading term in the energy density asymptotics. In fact, the scattering length will not appear but we will find the larger value $(8\pi)^{-1} \int v =(8\pi)^{-1} \widehat{v}(0)$ in its place. It should also be clear from the analysis above that we, indeed, did not ignore the quartic terms. 

The approximations above can be controlled rigorously \cite{NApiorkowskiReuversII}. The argument originally goes back to the paper \cite{ErdosSchleinYau}. In \cite{NApiorkowskiReuversII} it was shown by a more detailed analysis that minimizing the Bogolyubov functional in (\ref{eq:bofunctional}) does not give the second correction in the Lee-Huang-Yang formula. It does give the formula if $\widehat{v}(0)$ is replaced everywhere by the smaller quantity $8\pi a$. In fact, the difficulty in the rigorous analysis is exactly to understand this replacement. 

Restricting to quasi-free states, indeed, ignores the cubic terms of the form 
$$
\widehat{v}(k)a^\dagger_{k}a^\dagger_{q-k}a_qa_0 +\text{hermitian conjugates}
$$
in the Hamiltonian just as in the usual Bogolyubov approximation. It was, however, understood both in \cite{YauYin} and in \cite{Fournais2020,Fournais2022} that these terms cannot be ignored if the goal is to get the Lee-Huang-Yang order correctly. 
 
\section{Bose-Einstein Condensation}\label{sec:BEC}

In the previous section we saw that the optimal Bogolyubov trial state has almost complete condensation. It is, however, one of the major open problems in the mathematical analysis of Bose gases to show Bose-Einstein condensation (BEC) for the true ground state. The precise meaning of BEC is that in the thermodynamic limit we have for the true ground state $\Psi_L$ (indicating explicitly that it of course depends on $L$) that
$$
\lim_{\substack{L\to\infty\\N/L^d\to\rho}}L^{-3}\langle\Psi_L,a^\dagger_0a_0\Psi_L\rangle=\rho_0\ne0
$$
i.e., that we have macroscopic occupation in the zero momentum state. 

It turns out, however, that to rigorously establish the Lee-Huang-Yang formula we do not need to prove  BEC in the thermodynamic limit. It is enough to establish condensation for the gas confined to a finite size corresponding to the Gross-Pitaevskii regime. More precisely, in dimension $3$ this means that the gas shows condensation if confined to a box of length scale $L$ of order the {\it healing length} $\sim(\rho a)^{-1/2}$. On this scale condensation can be proved if the ground state energy can be approximated to sufficient accuracy. To understand this heuristically note that if $\rho_+=\rho-\rho_0$ is the density of particles not in zero momentum then in a cube of length $L$ these particles would have a kinetic energy density of at least 
$\rho_+ \pi^2 L^{-2}$. If we control the energy to the Lee-Huang-Yang order this would imply that 
$
\rho_+  L^{-2}\ll \rho^2 a\sqrt{\rho a^3}
$,
or 
$$\frac{\rho_+}{\rho}\ll (L \sqrt{\rho a})^2\sqrt{\rho a^3}.
$$ 

This was first seen rigorously for trapped potentials in \cite{Lieb2002} and improved in \cite{boccatoetal2018, boccatoetal2019, boccatoetal2020, adhikarietal2021, Fournais2021}. In the thermodynamic limit, however, condensation cannot be derived from controlling the energy density. The only known cases in which BEC has been established in the thermodynamic limit is for the hard core Bose gas on a lattice at half filling \cite{Kennedy1988,Aizenman2004}. 

It is an interesting question whether the experimental observations of BEC going back to the seminal works \cite{CornellWieman1995, Ketterle1995} can be said to be in a thermodynamic regime in the sense that the confinement is in a region large compared to the healing length, properly defined. The original works were in a harmonic trapping potential whereas more recent experiments\cite{Gaunt2013} are closer to the uniform case. The experiment considers confinement of ${}^{87}$Rb atoms in a cylindrical ``box'' with cross-sectional diameter  $R=35\mu m$ and length $L=70\mu m$. The experiment starts with $6\times 10^5$ particles before evaporative cooling and ends with a condensate density of $2x10^{12} cm^{-3}=2\mu m^{-3}$ corresponding to approximately $5x10^5$ condensate particles. It may not be entirely clear how to define the healing length in order to make a comparison. A reasonable comparison would be whether the leading order energy per particle, i.e $4\pi\rho a$, corresponds to many transversal kinetic energy modes. The number of transversal energy modes up to level $\lambda=4\pi\rho a$ may be estimated by the phase-space volume. With a scattering length of ${}^{87}$Rb around $5 \times 10^{-3}\mu m$ \cite{Burke1999} and using for $\rho$ the condensate density $2\mu m^{-3}$ we find for this estimate on the number of transversal modes 
$$
(2\pi)^{-2}\iint_{p^2\leq\lambda, r<R}d^2p d^2r=\frac14\lambda R^2=\pi \rho a R^2\approx35^2\pi/100\approx 38.
$$
This is a reasonably large number of modes. It would still be interesting to increase the size of the sample to push our understanding of BEC further. 

Likewise, it may be considered whether the experimental study of the Lee-Huang-Yang correction for ${}^7$Li atoms in \cite{Navon2011} is in a trapped (Gross-Pitaevskii regime) or closer to a thermodynamic regime. In this setup the healing length is approximately $1\mu m$ whereas the cylindrical trap (in this case harmonic) has cross-sectional radius $5\mu m$ and length $100\mu m$.

\section{Ideas behind the proof of the Lee-Huang-Yang formula}\label{sec:proof}

We will give a very sketchy overview of how to prove the lower bound in the Lee-Huang-Yang formula. The long and technical details can be found in \cite{Fournais2020} and \cite{Fournais2022}. There are two main steps in the proof. The first step is to control condensation by localizing the gas to a box of size of the healing length $\ell\sim(\rho a)^{-1/2}$ without paying too high a localization prize. 

In a box of that size we may conclude condensation as described above. In fact, if we only want to show $\rho_+\ll \rho$ it is enough to control the energy to any order better than the leading order $\rho^2 a$. 

In a box $B=[0,\ell]^3$ we consider the projection $P$ onto constant functions, i.e., 
$$
P=\ell^{-3}|1_B\rangle\langle1_B|
$$
and the orthogonal projection $Q=1-P$. Note that the operator $\sum_i P_i$ acting on the many-body space counts the number of particles in the condensate, i.e., $\sum_i P_i=a_0^\dagger a_0$ in second quantization, and $\sum_i Q_i$  counts the particles {\it not} in the condensate. Thus we know a-priori from the first step that $\sum_i Q_i$ is not too big. 

We may  write 
\begin{align*}
\sum_{i<j} v(x_i-x_j)=&\sum_{i<j} (P_i+Q_i)(P_j+Q_j)(v(x_i-x_j)(P_i+Q_i)(P_j+Q_j)\\=&\sum_{i<j} P_iP_jv(x_i-x_j)P_iP_j+\cdots +Q_iQ_jv(x_i-x_j)Q_iQ_j,
\end{align*}
where in the last expression we have expanded in 16 terms.
The usual Bogolyubov approximation corresponds to keeping terms with 2 or fewer $Q$ operators and ignoring the terms with 3 and 4 $Q$'s. We know however that this will not work. 

A crucial idea in \cite{Fournais2020} was to recognize that the expression that we can afford to ignore in the lower bound is the positive expression (using that $v \geq0$ )
$$
\sum_{i<j} (Q_iQ_j+(P_iP_j+P_iQ_j+P_jQ_i)\omega )v
(Q_iQ_j+\omega (P_iP_j+P_iQ_j+P_jQ_i))\geq0.
$$
Recall that $u=1-\omega$ was the scattering solution. The dependence of $\omega$ and $v$ on $x_i-x_j$ is omitted to lighten the notation.
Put differently we subtract this term from the Hamiltonian to get a lower bound. This will have the effect to remove the 4-$Q$ term and in the other terms replace $v$ by either $v(1-\omega)$ or $v(1-\omega^2)$. Note that replacing $v$ by $v(1-\omega)=vu$ gives us a potential (see (\ref{eq:u2a})) with the property that $\int vu=8\pi a$, which is the main point in this idea. 

There are still the $(1-\omega^2)v=(1-\omega)v+\omega(1-\omega)v$ terms as well as the 3-$Q$ terms. The 3-$Q$ terms require a difficult analysis before a standard Bogolyubov diagonalization can be performed. It leads to a calculation where the additional terms with $\omega(1-\omega)v$ exactly cancel to arrive at the Lee-Huang-Yang formula.

\section{Some reflections on the one-dimensional case}\label{sec:1D}

We will not discuss the  one-dimensional case of 
Theorem~\ref{thm:main} in details but will only make a few remarks about it. In the one-dimensional case there are two cases in which the ground state energy can, at least in principle, be determined explicitly. 

The simplest case is that of a hard-core potential. In this case the ground state energy of the Bose gas is identical to the ground state energy of the free Fermi gas in the smaller length ("volume") $L-Na$ corresponding to the higher density $\rho(1-\rho a)^{-1}$. Recalling that the ground state energy of the free Fermi gas in one-dimension is $\frac{\pi^2}3 N^3 L^{-2}$ an easy calculation shows that the ground state energy of the hard-core Bose gas is 
$
\frac{\pi^2}3 N\rho^2(1-\rho a)^{-2}
$. I.e., the ground state energy density is 
$$
e(\rho)=\frac{\pi^2}3 \rho^3 (1-\rho a)^{-2}=
\frac{\pi^2}3 \rho^3 (1+2\rho a+ o(\rho a)).
$$
In agreement with Theorem~\ref{thm:main}. 

The other case that can be treated exactly is the Lieb-Liniger gas \cite{Lieb1963} with delta interactions $v(x)=2c\delta(x)$, $c>0$. It was solved exactly in \cite{Lieb1963}. To be more precise calculating the ground state energy is reduced to solving an integral equation. This is not entirely trivial and a closed form expression does not exist. It is, however, known that in the dilute limit $\rho$ small or equivalently $c$ large (the strongly interacting case) we have
$$
e(\rho)= \frac{\pi^2}3 \rho^3 \left((1+2\rho/c)^{-2} + o(\rho/c)\right).
$$
Since the scattering length of $v$ in this case is negative with $a=-2/c$ we again arrive at the formula in Theorem~\ref{thm:main}. The rigorous proof that this is universally true for all potentials is, at least, for the lower bound based on a comparison with the Lieb-Liniger model. The details of both the upper and lower bound are given in \cite{agerskov2024}.

\section{Conclusion and summary of open problems}\label{sec:conclusion}
We have briefly discussed the known mathematical results on the ground state energy density of Bose gases in the dilute limit. For positive (repulsive) potentials we know under fairly general circumstances a two-term asymptotic expansion for the ground state energy density. The most important unsolved case is the hard-core potential in three dimensions. In this case, we know that the Lee-Huang-Yang formula gives a correct lower bound on the ground-state energy density. For the upper bound, the best result \cite{Bastietal24} is an estimate with a coefficient on the term of Lee-Huang-Yang order that is too large. Finally, it is still a major challenge to go beyond the Lee-Huang-Yang order and prove the Wu corrections in \eqref{eq:Wu} and the corresponding two-dimensional Mora-Castin  corrections in \eqref{eq:Moracastin}.

Another important open problem is generalizing to potentials that are allowed to be negative. The best result here is \cite{Yin-negV}. Of course, it must be a requirement that the potential has no $k$-body bound states for any finite particle number $k$. Thus, in particular, it rules out potentials with range smaller than the scattering length where there are two-body bound states. Thus, the situation studied in experiments with cold atomic gases where the scattering length is significantly larger than the range are at best meta-stable. Analyzing such systems mathematically would be a major challenge, as we would not have a true ground state. 

The most challenging mathematical problem for dilute gases is to establish Bose-Einstein condensation. It has been done for confined gases, but, except for a few cases that are certainly not dilute, it is an open problem in the thermodynamic limit corresponding to gases confined to regions much larger than the healing length. It might also be interesting to consider experiments that probe condensation even further into this regime.

\bibliographystyle{crunsrt}

%This calls all references from the .bib
\nocite{*}

%  This inserts the bib file
\bibliography{JPS}

\end{document}